# Barrier-Free Tunneling in a Carbon Heterojunction Transistor

Youngki Yoon[a)] and Sayeef Salahuddin[b)]

Department of Electrical Engineering and Computer Sciences,

University of California, Berkeley, CA 94720

**ABSTRACT**

Recently it has been experimentally shown that a graphene nanoribbon (GNR) can be obtained by unzipping a carbon nanotube (CNT). This makes it possible to fabricate all-carbon heterostructures that have a unique interface between a CNT and a GNR. Here we demonstrate that such a heterojunction may be utilized to obtain a unique transistor operation. By performing a self-consistent non-equilibrium Green's function (NEGF) based calculation on an atomistically defined structure, we show that such a transistor may reduce energy dissipation below the classical limit while not compromising speed – thus providing an alternate route towards ultra low-power, high-performance carbon-heterostructure electronics.

E-mail: [a)]yyoon@eecs.berkeley.edu, [b)]sayeef@eecs.berkeley.edu

Conventional transistors are thermally activated. A barrier is created whose height is modulated to control current flow. This modulation of the barrier changes the number of electrons exponentially following the Boltzmann factor $\exp(qV/k_BT)$. This in turn means that to change the current by one order of magnitude (or one decade) a voltage of at least $2.3k_BT/q$ (60 mV at room temperature) is necessary. This leads to the fact that the rate of change in voltage with respect to current on a log scale, often termed as subthreshold swing, has a classical limit of 60 mV/decade. Supply voltage requirement in a conventional transistor is constrained by this limit, currently leading to a situation that the future of electronics will be severely limited by the large voltage requirement and ensuing energy dissipation.[1]

It has been proposed that the classical constraint on supply voltage can be avoided by resorting to inter-band quantum tunneling so that the current flow is modulated by a tunneling barrier instead of a thermal barrier.[2,3] However, conventional tunneling field-effect transistors (TFETs) suffer seriously from the very presence of a tunneling barrier as it significantly limits the current flow and thereby the speed at which the transistor can be operated.

In this paper, we show that by utilizing a heterojunction created at the interface of a graphene nanoribbon (GNR)[4,5] and a metallic carbon nanotube (CNT)[6], a unique transistor structure may be possible, where current is not dictated by the Boltzmann factor, yet the tunneling barrier is greatly diminished. The basic idea is illustrated in Fig. 1(a). In a conventional TFET, the current must tunnel through the barrier (shown by the dashed line), the amplitude of the current being severely limited by the significant resistance posed by the barrier. One could, on the other hand, imagine that if the tunneling barrier is replaced by a material with zero bandgap (shown by the shaded region), a large current should flow unhampered due to the absence of any tunneling barrier. However, to obtain this situation in practice, a very unique combination of materials is needed: (i) a material with zero bandgap whose band alignment could be modulated by a gate (which means ordinary metals would not work), and (ii) good wavefunction overlap of the zero bandgap material with left p and right i (intrinsic) regions to ensure a large current. Both these conditions can be satisfied with a heterostructure of CNT and GNR obtained by partially unzipping a metallic CNT. Notably, both the requirements, namely (i) the control of bands in a metallic CNT by doping or gate voltage[2,7] and (ii) unzipping a CNT to produce a CNT-GNR interface[4,5], have already been experimentally demonstrated.

Here we consider a single-walled (6,0) CNT that has been predicted to be metallic by both a simple $p_z$ tight-binding model[8] and *ab-initio* studies[9]. It is energetically stable and CNTs of even smaller diameters have already been experimentally deomonstrated[10]. The two ends of the CNT are connected to semiconducting *n* = 12 armchair-edge GNR (a-GNR) unzipped from the (6,0) CNT, assuming no loss of carbon atoms [Fig. 1(b)]. The schematic of a transistor made out of this interface is shown in Fig. 1(c). We shall call this a carbon-based heterojunction FET (HFET), which eventually combines the concepts of tunneling FET and conventional FET in the same device structure. Note that any GNR partially unzipped from a reasonably small-sized metallic CNT is expected to be semiconducting in nature (See supplementary material). The *n* = 12 a-GNR is assumed to be hydrogen-terminated, having a bandgap of $E_g$ = 0.6 eV.

Once the atomistic structure is constructed, device characteristics are calculated from a self-consistent solution of 3-D Poisson and non-equilibrium Green's function (NEGF) based transport with atomistic resolution. A nearest-neighbor tight-binding approximation was used with a $p_z$ orbital basis set ($t_0 = -2.7$ eV) in real space approach (See supplementary material). An infinite-potential-well boundary condition was used along the width of the semiconducting $n = 12$ a-GNR, whereas a periodic boundary condition was applied along the circumferential direction of the metallic (6,0) CNT (See supplementary material for details of the Hamiltonian matrix and the equations solved). A modified tight-binding parameter ($t = 1.12\, t_0$) was used for the edges of GNR to treat edge-bond relaxation.[11] The effect of doping was treated by the number of dopants per each carbon atom ($N_{S/D}$, $N_{CNT}$). In practice, GNR doping can be achieved by chemically functionalizing the edges.[12]

Figure 2 summarizes the simulated results. Transfer and output characteristics are shown in Figs. 2(a) and 2(b), respectively. Notably, near the threshold voltage $V_t$ ($\approx 0.15$ V), which is defined where a significant current starts to show up in the linear scale, the current changes by more than 4 orders of magnitude with a change in voltage of only 100 mV, thus providing a subthreshold swing that is much below the classical limit of 60 mV/decade [Fig. 2(a)]. To compare the ON current, we also simulated a TFET based on all-semiconducting GNR with other device parameters such as source/drain doping density, gate geometry, gate oxide, source/drain extension, and power supply voltage unchanged (See supplementary material). An HFET can deliver up to ~70 times larger current than a TFET at the ON state ($V_G = V_t + 0.15$). Since no bandgap exists inside the metallic CNT region of the HFET [Fig. 2(c)], the carrier transport takes place through the gapless energy states [Fig. 2(d)] of the p-n junction created inside the metallic CNT [left panel in Fig. 2(g)], where the transmission probability can be as large as unity due to the absence of back scattering stemming from the linear energy dispersion relation near the Fermi level.[6] The maximum value of the simulated transmission probability of the HFET is 0.9978 [Fig. 2(h)], which is very close to the theoretical maximum value of 1. By contrast, the maximum transmission probability is 0.015 for a GNR TFET, where a current flow is fundamentally limited by the tunneling barrier posed by the bandgap of GNR (Fig. S1). Therefore, the simulated characteristics show that the HFET can (i) provide a subthreshold swing less than the classical limit and (ii) at the same time efficiently eliminate the tunneling barrier so that the drive current can be significantly large.

Looking at Fig. 2(a), one would see an asymmetry in the current of the HFET with respect to the gate voltage. To understand this, we plotted the energy-resolved current spectrum in Figs. 2(d), 2(e) and 2(f) at positive, zero and negative gate voltages, respectively. At $V_G \geq 0$, electrons are injected near the source Fermi level [Figs. 2(d) and 2(e)]. For $V_G < 0$, however, the spectrum of current exists around the drain Fermi level, which indicates hole flow [Fig. 2(f)]. It is clear from these observations that the mechanism of current flow is fundamentally different at negative gate voltages from that at positive gate voltages. At a large positive gate voltage, current flows through the 'barrier-free' channel [Fig. 2(d)], whereas for negative voltages, the current must tunnel through a barrier by inter-band tunneling [Fig. 2(f)]. In addition, the use of a partial gate further adds to the suppression of current at negative voltages. As a result, at $V_G = -0.2$ V, the current is still 3 orders of magnitude smaller than the current at a similar positive $V_G$ [Fig. 2(a)].

Ordinarily, a band bending will exist at the CNT-GNR interface near the source ($x = 15$ nm) due to the misalignment of Fermi levels of the two regions. For the simulated device structure, we have used a large doping ($N_{CNT} = 10\ N_{S/D}$) partially inside the metallic CNT to show how this equilibrium band bending can be minimized (Fig. 3). In practice, electrostatic doping by a separate gate is widely used to emulate the effect of local doping.[2] Note that if significant band bending remains, the carriers must tunnel through a barrier at the source/channel interface and hence the current density will be limited [Fig. 3(b)]. If an ordinary metal were used, such reduction of band bending would not be possible, since the bands inside the metal cannot be effectively controlled by electrostatic means. As for the junction at the other side ($x = 35$ nm), there is no equilibrium band bending since it is a junction of intrinsic semi-metal and intrinsic semiconductor. The small tip existing at this interface is caused by the gate bias as observed at regular Schottky contacts. This has minimal influence on the current as the size of the barrier is very small.

By artificially changing the curvature of the carbon heterostructure, we have verified that such variation has a minimal effect on electrostatics (See supplementary material for details: Fig. S2). However, edge roughness can result in bandgap variations in the GNR region and thus degrade performance. It is expected, though, that in the unzipped GNR, the edge roughness would be significantly reduced.[4,5] Furthermore, various scattering phenomena and surface morphology may also adversely affect the transistor performance.

To summarize, we have shown that it may be possible to use the CNT-GNR heterostructure to obtain a unique transistor action such that (i) at low voltages the device acts like a tunneling transistor, reducing the voltage requirement below the classical limit and (ii) at high voltages the tunnel barrier is effectively diminished, thus allowing a large ON current. Therefore, such a heterostructure may make it possible to combine the best practices of both the tunneling FET and the conventional FET in the same device structure and thus guide the way in designing ultra low-power carbon-heterostructure electronics.

The authors thank Cheol-Hwan Park for helpful discussions. This work was supported in part by FCRP center on Functional Engineered and Nano Architectonics (FENA).

# FIGURE CAPTIONS

**Figure 1. Carbon-based heterojunction field-effect transistor (HFET).**

**(a)** Schematic band structure at ON (red solid line) and OFF states (blue solid line). Proposed device has a semi-metallic zero-bandgap region (shaded). $E_{FS}$ and $E_{FD}$ are Fermi levels at the source and the drain, respectively. The dashed lines show the bandgap of normal p-i-n structure in the absence of semi-metal at ON state.

**(b)** Atomistic configuration of a carbon nanotube (CNT) and graphene nanoribbon (GNR) heterostructure.

**(c)** Schematic device structure. The channel is a carbon-based heterostructure that consists of a metallic (6,0) CNT and semiconducting $n = 12$ armchair-edge GNRs (a-GNRs). Double-gate geometry with 1.5 nm thick $HfO_2$ gate oxide ($\kappa = 16$). Lengths of source/drain extensions, metallic CNT, gate-controlled channel, and unbiased channel are $L_{S/D} = 15$, $L_{CNT} = 20$, $L_G = 20$, $L_U = 15$ nm, respectively. Source (drain) GNR is p-doped (n-doped) with doping density, $N_{S/D} = 1.8 \times 10^{-3}$ /atom, which is equivalent to 0.1 /nm. Effective doping density at the left half of the metallic CNT is $N_{CNT} = 10\, N_{S/D}$. Power supply voltage is $V_{DD} = 0.4$ V.

**Figure 2.**

**(a)** $I_D - V_G$ characteristics on a log scale (blue curve with left axis) and on a linear scale (green curve with right axis). Red dots indicate several important points such as threshold voltage (c), ON state (d), and OFF states (e, f). Subthreshold swing, $S = dV_G/d(\log_{10} I_D)$ are shown for various gate bias ranges. It shows an ambipolar conduction by both electrons and holes. Current saturates beyond $V_G = 0.3$ V as a potential hump appears [see the arrow-indicated region in Fig. 2(d)] due to the partial gate structure. (See supplementary material for details: Fig. S2). The minimum leakage current is shifted to $V_G = 0$ by using appropriate metal work function engineering ($\Phi_{ms} = 0.15$ eV).

**(b)** $I_D - V_D$ plot. The output characteristic is analogous to that of an ordinary transistor except in the low voltage where a tunneling behavior can be clearly observed.

**(c)** Local density-of-states (LDOS), shown on a log scale, at threshold voltage, $V_t = 0.15$ V. Metallic CNT does not have bandgap and states exist at the entire energy levels. The solid lines show the bandgap of the semiconducting GNR. The dashed line shows the self-consistent electrostatic solution of where the conduction band and the valence band touch each other within the metallic region.

**(d)** Current spectrum (red strip) at ON state ($V_G = 0.3$ V).

**(e-f)** Current spectrums (red strips) at OFF states. The different energy levels of the current spectrums indicate a transition from the electron conduction (e, $V_G = 0$ V) to the hole conduction (f, $V_G = -0.05$ V).

**(g)** Zoom-in plot of the spatial distribution of current as a function of energy for the region shown inside the box of Fig. 2d (left). The right panel shows energy-resolved current density. The sharp feature appearing at $E \approx -0.11$ eV is due to the resonant tunneling states originating from the barrier profile at this specific voltage.

**(h)** Transmission probability, $T$ as a function of energy, $E$ and gate voltage, $V_G$. The largest value of $T$ is 0.9978.

**Figure 3. Effect of doping in the metallic CNT.**

**(a)** Band profiles for CNT doping density, $N_{CNT} = N_{S/D}$ (insufficient doping, red dashed lines) and $N_{CNT} = 10\ N_{S/D}$ (used in our simulation, blue solid lines). The left half of the metallic CNT is doped to reduce the inherent band bending at the heterojunction. Red strip shows current spectrum for $N_{CNT} = N_{S/D}$ at $V_G = 0.3$ V.

**(b)** Energy-resolved current density at $V_G = 0.3$ V.

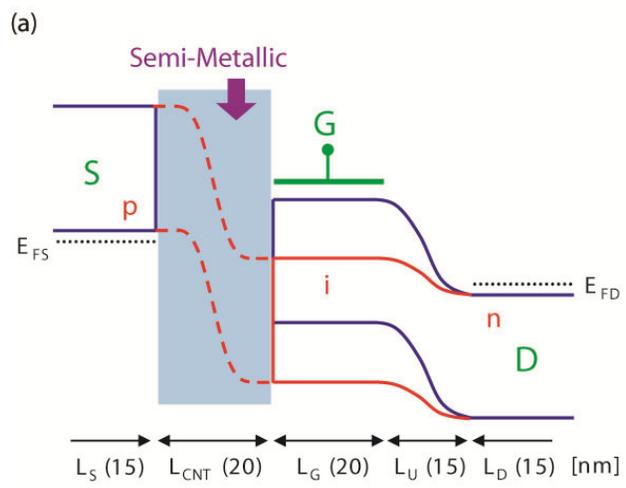 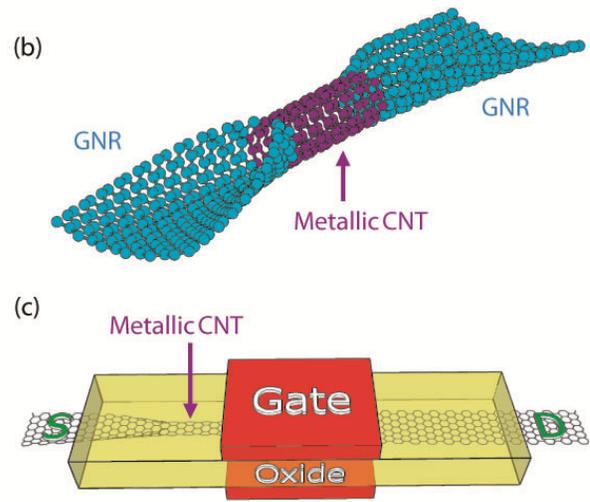

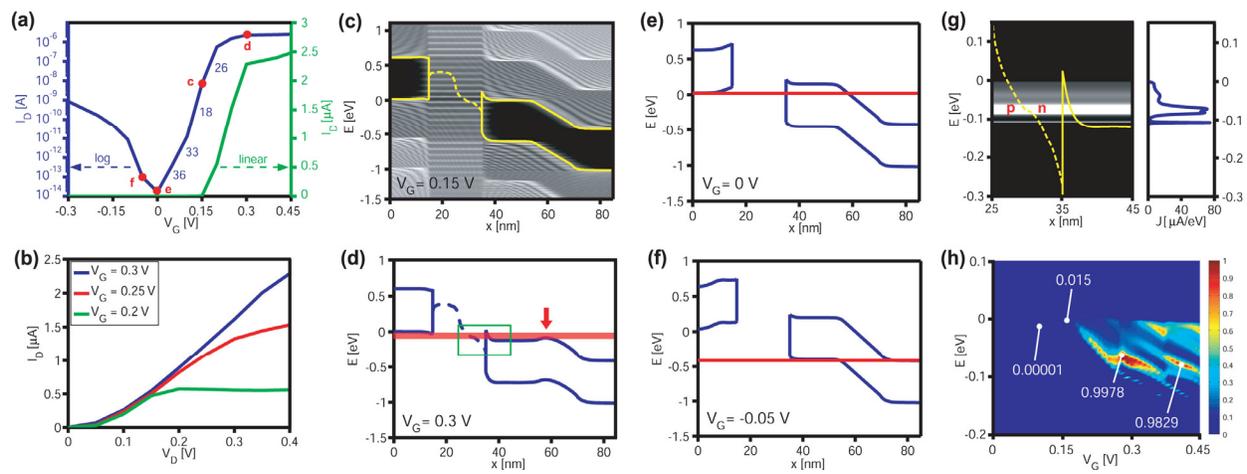

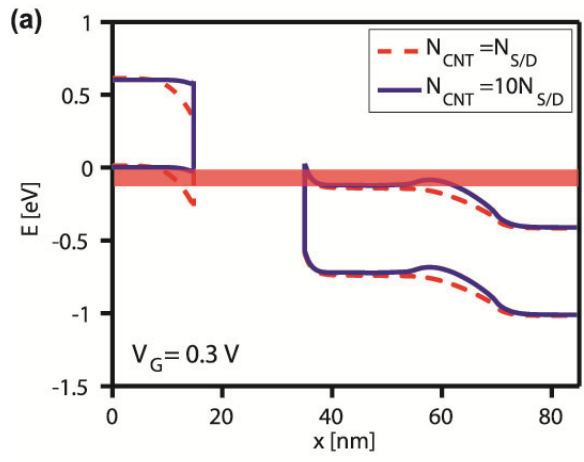 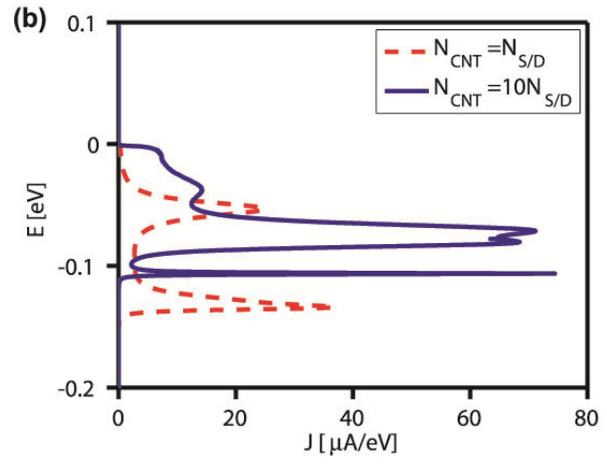